\documentclass[]{pasj01} 
\usepackage{graphicx}	
\usepackage{bm}  

\title{One-kpc Expanding Cylinder of HI-Gas around the Galactic Center}

\author{Yoshiaki SOFUE\\
Institute of Astronomy, The University of Tokyo, Mitaka, Tokyo 181-0015, Japan  (Email: sofue@ioa.s.u-tokyo.ac.jp)}

\date{2022} 
  
\def\vlsr{V_{\rm LSR}}

\def\Msun{M_\odot} 
\def\deg{^\circ} \def\Tb{T_{\rm b}}
   
\def\be{\begin{equation}} \def\ee{\end{equation}}
\def\red{\textcolor{red}}   
 
\def\Xhi{X_{\rm HI}} 
\def\kms{km s$^{-1}$}
\def\ekms{{\rm ~km~s^{-1}~}} 
 \def\rmh2{{\rm H_2}}

\def\red{\textcolor{red}}
\def\red{}

\KeyWords{ISM: atoms --- ISM: jets and outflows --- Galaxy: center --- Galaxy: kinematics and dynamics}
\begin{document} 

\maketitle  
%%%%%%%%%%%%%%%%%%%%%%%%%%%%%%%%%%%%%%%%%%%%%%%%%%%%%%%%%     
\begin{abstract} 
We report the discovery of an expanding cylinder of HI gas of radius 1 kpc and vertical extent 800 pc by analyzing the 21-cm line survey data from the literature.
The cylinder is expanding at 150 \kms and rotating at 100 \kms, and is interpreted as due to a high-velocity conical wind at $\sim 180$ \kms from the Galactic Center.
The total mass of the cylinder is estimated to be $\sim 8.5\times 10^5 \Msun$ and kinetic energy $\sim 3\times 10^{53}$ ergs.  
\end{abstract} 

%\linenumbers

\section{Introduction}

Expanding rotating rings and cylinders concentric to the Galactic Center (GC) have been observed at various radii in the Galactic disc in the radio line emissions, exhibiting tilted ellipses in longitude-radial velocity diagrams (LVD).
The 3-kpc and 2.4-kpc expanding rings were discovered as tilted LV ellipses by the early HI line observations  \citep{sanders+1972,sanders+1973,simonson+1973}. 
A cavity in the halo having conical boundary has been found in the HI halo indicating a galactic wind from the GC  
\citep{lockman+2016,lockman+2020,sofue2017b,sofue2022}. 
The $200$-pc expanding molecular ring (EMR) has been known as the evidence for the GC explosive activity \citep{kaifu+1972,scoville1972,sofue1995a,sofue1995b,sofue2017a,krum+2017}. 
In these observational studies, the tilted LV ellipse has been interpreted as due to expansion of a ring or cylinder rotating around the GC.

On the other hand, dynamical studies of the Galactic disc with a bar potential have shown that the bar-induced non-circular motion of gas along the X orbits can also explain the tilted ellipses and parallelograms in the LVD \citep{binney+1991,sormani+2015,rodoriguez+2008}. 
It is, therefore, controversial whether the LV ellipses are due to expanding motion or non-circular flow in a bar potential.

In this paper, we present a new case of LV ellipse at high latitudes, which cannot be explained by a bar potential model, but is naturally attributed to an expanding gas flow from the GC.

\section{Tilted Ellipse in Longitude-Velocity Diagram}

\begin{figure} 
\begin{center}
\includegraphics[width=6.6cm]{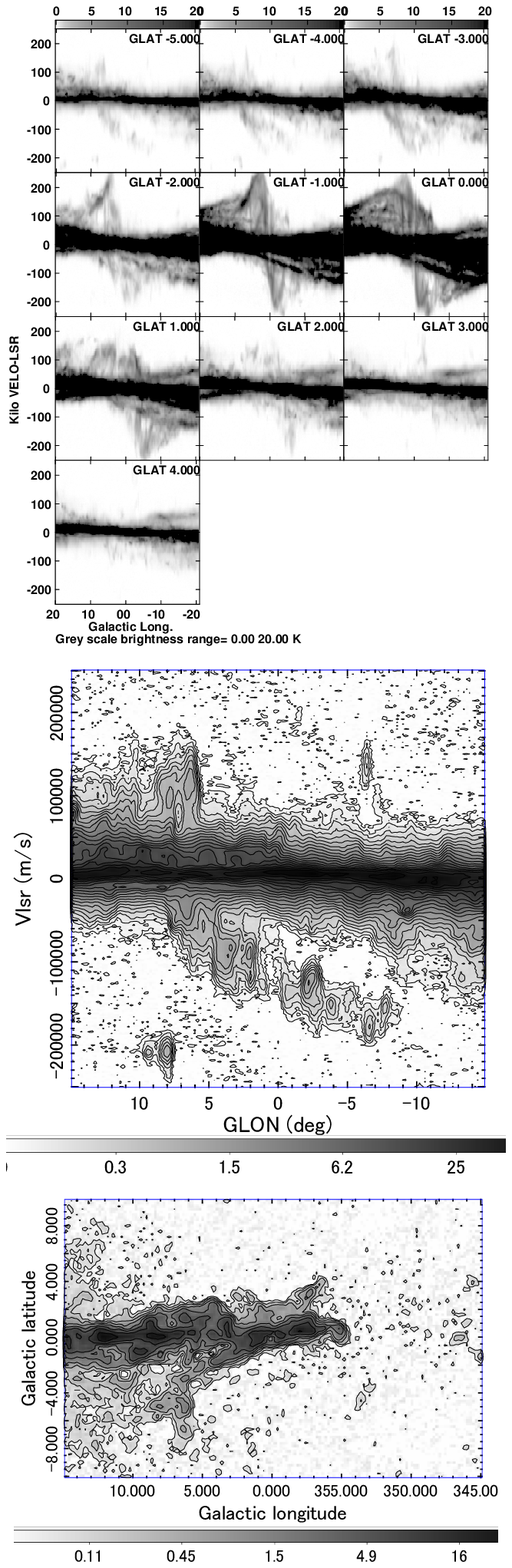} 
\end{center}
\caption{\red{(a) HI-line LVD of the central $40\deg \times 40\deg$ region, each averaged within every $1\deg$ latitudinal interval between $b=-5\deg$ and $-4\deg$ (top left) to $+4\deg$ and $+5\deg$ (bottom right).}
(b) Enlargement of LVD averaged between $b=-5\deg$ to $-4\deg$. 
(c) $\Tb$ channel map at $\vlsr=+120 \ekms$.}
\label{cub120} 
\end{figure}

\red{In figure \ref{cub120} (a) we show HI-line LVD in the central $40\deg\times 10\deg$ region,} each averaged within latitudinal interval of $1\deg$ in the central $l=\pm 20\deg$ region, as produced from the HI Galactic All Sky Survey (GASS) \citep{mcclure+2009,HI4PI2016}.
Besides the bright disc structures near the galactic plane and the concentrated GC disc, we notice the vertically extended, tilted and curved ridges at latitudes $|b|\ge \sim 2\deg$ at $l\sim \pm 5-10\deg$.
Although fainter than the disc structures, they are clearly visible with brightness of $\Tb\sim 1-5$ K, showing large non-circular motions.
The features are also visible in the higher-resolution HI survey of the central $\pm 10\deg$ region \citep{mcclure+2012}.
In the intensity channel maps, the HI disc near the tangent velocities at $\vlsr \sim \pm -100 - 200 \ekms$ exhibits vertically extended spurs, some of which correspond to the non-circular LV ridges.

Figure \ref{cub120}(b) enlarges the LVD at  $b=-4\deg.5$ ($h\sim -630$ pc), where a curved ridge composing a part of an LV ellipse clearly shows up, revealing high non-circular velocities.
Panel (c) of the figure shows a $\Tb$ channel map at $\vlsr=+120 \ekms$, representing the gas distribution on the sky at this velocity.
It reveals a obliquely extending spur toward the south east corresponding to the left-side edge of the LV ellipse in the top panel.

\begin{figure} 
\begin{center}
\includegraphics[width=8cm]{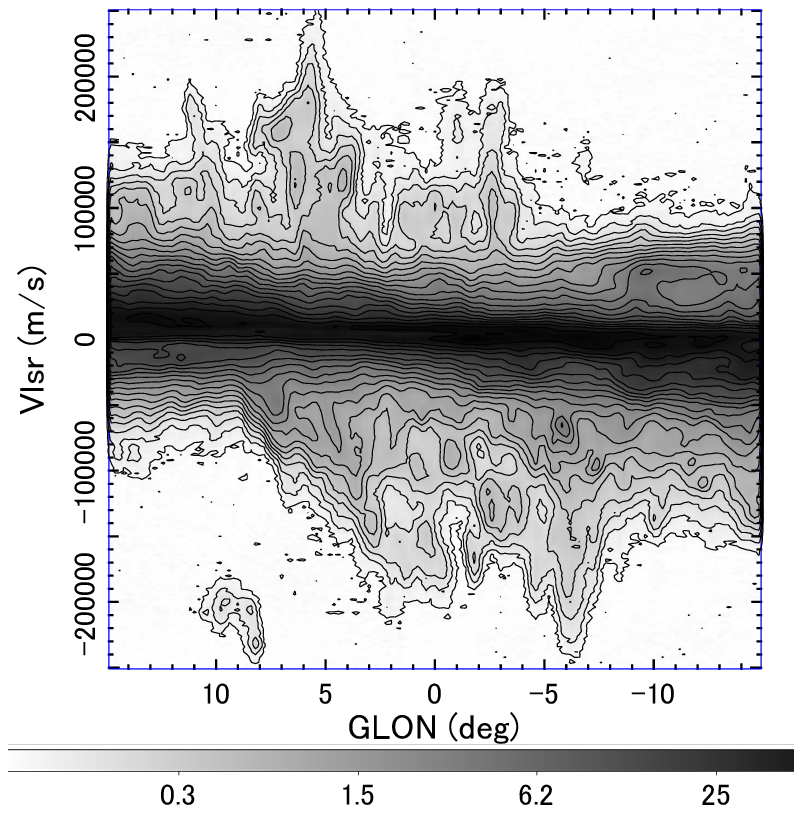}
\end{center}
\caption{HI LVD averaged in $2\deg \le |b| \le 4\deg$ ($|h|=280$ to 560 pc). }
\label{lvsum} 
\end{figure}

Figure \ref{lvsum} shows an LV diagram averaged at $2\deg \le |b| \le 4\deg$, or at vertical heights from $|h|=280$ to 560 pc.
The vertically extended, tilted and curved LV ridge shows up clearly here, composing a coherent tilted ellipse centered on the GC.
The maximum and minimum velocities along the ellipse are $\vlsr \sim \pm 250 \ekms$ and the longitudinal extent is $l\sim \pm 7\deg.5$ ($\pm 1.0$ kpc).
The intersections of the ellipse with the rotation axis of the Galaxy at $l=0\deg$ occur at $\vlsr\sim \pm 150 \ekms$, indicating that the ellipse is expanding at this velocity.
The tangential velocity at the maximum longitudes appears at $\vlsr \sim \pm 100 \ekms$, indicating that the ring is rotating at $\sim 100 \ekms$ in the same direction as that of the galactic rotation.
The HI intensity attains the maximum at the left and right side edges of the ellipse, and minimum or lacking region appears near the rotation axis.
\red{Another noticeable point is the isolation of the LV ellipse, uniquely standing alone surrounded by emission vacancy along the terminal-velocity envelope.}

\begin{figure} 
\begin{center}
\includegraphics[width=8cm]{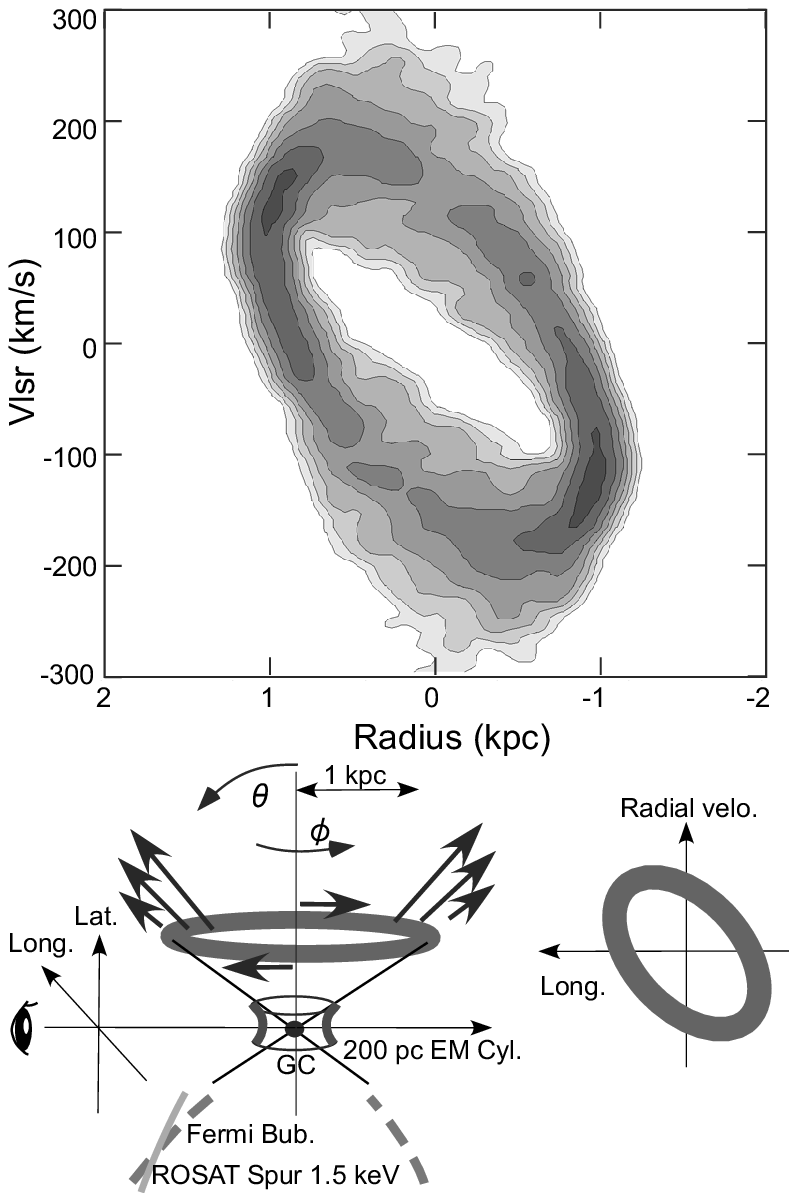} 
\end{center}
\caption{Simulated LVD of the conical cylinder wind of HI gas and illustration of the parameters. Note the difference from LVD of an expanding ring showing uniform brightness as illustrated in the bottom right. }
\label{model} 
\end{figure}

Using the diagram, we then estimate the mass of the ellipse, assuming that the distance to the GC is 8 kpc.
First we measure the mean brightness temperature $\Tb$ at velocities higher than $|\vlsr|\le 75 \ekms$ in order to avoid the galactic disc's emission.
By multiplying the velocity channel width and grid interval to the obtained total counts of $\Tb$, we calculate the mass of the measured area to be $M_{\rm measured}\sim 2.4 \times 10^5 \Msun$.
Here, we used a conversion factor of $\Xhi=1.823 \times 10^{18}$ H cm$^{-2}$ [K \kms]$^{-1}$ and the mean atomic weight including metals per hydrogen of $\mu=1.4$.
The thus estimated mass represents a value appearing in figure \ref{lvsum}, partially filling the whole structure by a factor of $\eta\sim 4\deg/10\deg \times (400-150)\ekms/400 \ekms \sim 0.32$.
So, we correct for the filling factor, and obtain the total gaseous mass composing the LV ellipse to be 
$M_{\rm tot}\sim M_{\rm measured}/\eta \sim 8.5 \times 10^5\Msun$.

\section{Discussion and summary}

The HI ellipse in the LV diagram can be traced up to latitudes as high as $b\sim \pm 5\deg$, or $h\sim \pm 800$ pc from the galactic plane.
\red{From the brightness temperature of $\Tb \sim 2$ K and velocity width $\delta v\sim 30 \ekms$, HI column density at this height is estimated to be $N_{\rm HI}\sim 10^{20}$ H cm$^{-2}$.
This yields volume density of $n_{\rm HI}\sim 0.1$ H cm$^{-3}$ for a supposed line-of-sight depth of $\sim 300$ pc.
The density is three orders of magnitudes higher than that of the HI galactic disc of $10^{-4}$ H cm$^{-3}$ and two orders of magnitudes higher than the warm HI disc of $10^{-3}$ H cm$^{-3}$ at the same height \citep{sofue2019}.
Even if the gas is galactic-shock compressed by a factor of ten, the density is higher than that of the disc or an arm by an order of magnitude or more.
This means that the LV ellipse is not an extension of arm and/or ring structures associated with the galactic disc.
Furthermore, the stellar bulge and bar density at this height is an order of magnitude lower than that in the disc \citep{dwek+1995}, which means that the bar potential there is too shallow to produce the observed non-circular velocities of $\sim 100 \ekms$. } 

\red{From these considerations, we may conclude that dynamical models incorporating bar-induced oval orbits of the disc gas may be excluded as an explanation of the observed non-circular motion in the high-latitude LV ellipse.
This conclusion is confirmed by a detailed comparison of figures 1 and 2 with the simulated LV patterns around $l\sim \pm 8\deg$ in the most extensive LVD study of bar-induced non-circular flows in the galactic plane by two dimensional treatment \citep{sormani+2015}. }

An alternate model to explain the non-circular LVD is an expanding ring or cylinder model of gaseous outflow driven by explosion and/or high-speed wind from the GC.
In the following, we stand on this model, and estimate the parameters of the expanding motion.
As estimated by fitting to the LV ellipse, the ring radius, expansion velocity, and rotation velocity are $\sim 1$ kpc, $\sim 150$ \kms, and $100$ \kms, respectively.
\red{In the expanding ring model, the slower rotation than the galactic rotation of $\sim 200$ \kms is attributed to the conservation of angular momentum, because the gas is accumulated from the inner region by wind or a shock wave from the GC.}
 
The observed LVD in figure \ref{lvsum} shows significant intensity variation along the ellipse in such a sense that the intensity attains maximum in the left and right side edges, and minimum, or even missing, around $l\sim 0\deg$.
\red{We recall that such non-uniform LV ellipse can be explained by an outflow with variable expanding velocity \citep{sofue2022}. 
We here examine a simple case of a varying flow velocity model, where the line-of-sight velocity is given by
\begin{equation}
V=V_0(1+\delta) e^{1-(r/r_0)^2}{\rm sin} \theta ~ {\rm sin} \phi + V_{\rm rot}{\rm cos} \phi,
\end{equation}
where $\delta$ is fractional velocity dispersion with respect to $V_0$, 
$r$ is the radius, $\theta$ is the opening angle of the wind from the $z$ axis, $\phi$ is the azimuth angle of the element around the rotation axis. Here, $V_0$ and $r_0$ are constants taken to be 100 \kms and 1 kpc, respectively, and $delta$ is taken to be 0.5.}
\red{The radial distribution of gas density is expressed by
\begin{equation}
\rho=\rho_0 e^{-[(r-r_0)/\Delta r]^2},
\end{equation}
where $\rho_0$ is the density at $r_0=1$ kpc and $\Delta r=0.2 r_0$ is its width.
From the height to radius ratio of the LV ellipse of $5\deg$ to $7\deg.5$, we calculate the opening angle of the cone from the rotation axis to be 
$\theta  \sim 56\deg$.
Then the measured velocity of $\vlsr=\pm 150 \ekms$ at $l\sim 0\deg$ yields a flow velocity to be $V_0\sim 180 \ekms$. }

Figure \ref{model} shows the calculated result with an illustration of the model.
The simulation well reproduces the observed characteristics of LV ellipse as follows:
\begin{itemize}
\item Tilted LV ellipse fitted by $V_{\rm flow}\sim 180 \ekms$ and $V_{\rm rot}\sim 100 \ekms$.
\item Intensity maximum on the left and right side edges of the ellipse.
\item Intensity minimum near $l=0\deg$.
\end{itemize}
However, the lopsidedness as observed in figure \ref{lvsum} is not reproduced by the present model, which may be attributed to localized density distributions and/or to the asymmetric wind from the GC.

From the adopted parameters, we can estimate the total kinetic energy of the expanding motion of the ellipse to be $E_{\rm expa}\sim 3\times 10^{53}$ ergs.
The time scale for the expanding gas to reach the radius of $\sim 1$ kpc from the GC is on the order of $t\sim r/V_{\rm flow}\sim 1 {\rm kpc}/180 \ekms \sim 6$ My.
These may be compared with the energy of $\sim 10^{55}$ ergs of the strongest GC activity associated with several to 10-kpc scale giant bubbles in radio, X-, and $\gamma$-rays with time scale of $\sim 10$ My \citep{kataoka+2018}.
The HI wind may be one of intermittent minor events among the outflows from the GC.

On the sky, the root of the HI cone roughly coincides with the ridges of the Fermi bubbles and an X-ray 'claw' \citep{su+2010,kataoka+2015}, and appears to be a continuation of the more central expanding molecular cone of 200 pc radius \citep{sofue2017b}, as illustrated in figure \ref{model}.
The present outflow model is consistent with the current wind models
%(Lockman et al. 2016, 2020; Bland-Hawthorn and Cohen 2003; Sofue 2017b, 2021),
\citep{bh+2003,mcclure+2013,lockman+2016,lockman+2020,sofue2017b,sofue+2021}, 
providing complimentary physical quantities such as the flow and rotation velocities and the energy.

%%%%%%%%%%%%%%%%%%%%%%%%%%%%%%%%%%%%
\vskip 5pt
\noindent{\bf Acknowledgements}:
The author expresses his grateful thanks to the GASS HI survey group.

\noindent{\bf Data availability}:
The HI-line data were downloaded from the url:
https: //www.atnf.csiro.au/ research/ GASS/ index.html
%https://www.atnf.csiro.au/research/HI/sgps/GalacticCenter/Home.html.

%%%%%%%%%%%%%%%%%%%%%%%%%%%%%%%%%%%%%%%%%%%%%%%%%


\begin{thebibliography}{}

\bibitem[Binney et al. (1991)]{binney+1991} Binney J., Gerhard O.~E., Stark A.~A., Bally J., Uchida K.~I., 1991, MNRAS, 252, 210  

\bibitem[Bland-Hawthorn \& Cohen(2003)]{bh+2003} Bland-Hawthorn, J. \& Cohen, M.\ 2003, \apj, 582, 246.% doi:10.1086/344573

\bibitem[Contopoulos \& Papayannopoulos(1980)]{contopoulos+1980} Contopoulos, G. \& Papayannopoulos, T.\ 1980, \aap, 92, 33%bar, X orbits 

\bibitem[Dwek et al.(1995)]{dwek+1995} Dwek, E., Arendt, R.~G., Hauser, M.~G., et al.\ 1995, \apj, 445, 716. %doi:10.1086/175734

\bibitem[HI4PI Collaboration et al.(2016)]{HI4PI2016} HI4PI Collaboration, Ben Bekhti, N., Floer, L., et al.\ 2016, \aap, 594, A116. %doi:10.1051/0004-6361/201629178


\bibitem[Kaifu et al.(1972)]{kaifu+1972} Kaifu, N., Kato, T., \& Iguchi, T.\ 1972, Nature Physical Science, 238, 105.% doi:10.1038/physci238105a0

\bibitem[Kataoka et al.(2015)]{kataoka+2015} Kataoka, J., Tahara, M., Totani, T., et al.\ 2015, \apj, 807, 77.% doi:10.1088/0004-637X/807/1/77

\bibitem[Kataoka et al.(2018)]{kataoka+2018} Kataoka, J., Sofue, Y., Inoue, Y., et al.\ 2018, Galaxies, 6, 27.% doi:10.3390/galaxies6010027

\bibitem[Krumholz et al.(2017)]{krum+2017} Krumholz, M.~R., Kruijssen, J.~M.~D., \& Crocker, R.~M.\ 2017, \mnras, 466, 1213. %doi:10.1093/mnras/stw3195

\bibitem[Lockman \& McClure-Griffiths(2016)]{lockman+2016} Lockman, F.~J. \& McClure-Griffiths, N.~M.\ 2016, \apj, 826, 215.% doi:10.3847/0004-637X/826/2/215 HI wind from GC
\bibitem[Lockman et al.(2020)]{lockman+2020} Lockman, F.~J., Di Teodoro, E.~M., \& McClure-Griffiths, N.~M.\ 2020, \apj, 888, 51L.% doi:10.3847/1538-4357/ab55d8 HI accretion FB

\bibitem[McClure-Griffiths et al.(2009)]{mcclure+2009} McClure-Griffiths N.~M., et al., 2009, ApJS, 181, 398 %GASS 
\bibitem[McClure-Griffiths et al.(2012)]{mcclure+2012} McClure-Griffiths, N.~M., Dickey, J.~M., Gaensler, B.~M., et al.\ 2012, \apjs, 199, 12. %doi:10.1088/0067-0049/199/1/12%ATCA HI GC
\bibitem[McClure-Griffiths et al.(2013)]{mcclure+2013} McClure-Griffiths, N.~M., Green, J.~A., Hill, A.~S., et al.\ 2013, \apjl, 770, L4. %doi:10.1088/2041-8205/770/1/L4 %Hi-velo outflow clumps


\bibitem[Rodriguez-Fernandez \& Combes(2008)]{rodoriguez+2008} Rodriguez-Fernandez, N.~J. \& Combes, F.\ 2008, \aap, 489, 115.% doi:10.1051/0004-6361:200809644

\bibitem[Sanders \& Wrixon(1972)]{sanders+1972} Sanders, R.~H. \& Wrixon, G.~T.\ 1972, \aap, 18, 92
\bibitem[Sanders \& Wrixon(1973)]{sanders+1973} Sanders, R.~H. \& Wrixon, G.~T.\ 1973, \aap, 26, 365 %HI GC disc to 2.4, 3kpc expa. arms

\bibitem[Scoville(1972)]{scoville1972} Scoville, N.~Z.\ 1972, \apjl, 175, L127 

\bibitem[Simonson \& Mader(1973)]{simonson+1973} Simonson, S.~C. \& Mader, G.~L.\ 1973, \aap, 27, 337 %HI expanding arms
 
\bibitem[Sofue(1995a)]{sofue1995a} Sofue, Y.\ 1995a, \pasj, 47, 527 %GcA I
\bibitem[Sofue(1995b)]{sofue1995b} Sofue, Y.\ 1995b, \pasj, 47, 551 %GCA II
\bibitem[Sofue(2013)]{sofue2013} Sofue, Y.\ 2013, \pasj, 65, 118. %doi:10.1093/pasj/65.6.118%GC RC
\bibitem[Sofue(2017a)]{sofue2017a} Sofue, Y.\ 2017a, \mnras, 470, 1982. %doi:10.1093/mnras/stx1389%GC 3D EMcone
\bibitem[Sofue(2017b)]{sofue2017b} Sofue, Y.\ 2017b, \pasj, 69, L8.% 3kpc crator doi:10.1093/pasj/psx067
\bibitem[Sofue(2019)]{sofue2019} Sofue, Y.\ 2019, \mnras, 484, 2954. %doi:10.1093/mnras/stz143
\bibitem[Sofue(2022)]{sofue2022} Sofue, Y.\ 2022, \mnras, 509, 5809. % M17 bipolar lobe, doi:10.1093/mnras/stab3091
\bibitem[Sofue \& Kataoka(2021)]{sofue+2021} Sofue, Y. \& Kataoka, J.\ 2021, \mnras, 506, 2170.% doi:10.1093/mnras/stab1857


%\bibitem[Sormani et al.(2015)]{sormani+2015III} Sormani, M.~C., Binney, J., \& Magorrian, J.\ 2015, \mnras, 454, 1818. 
\bibitem[Sormani \& Magorrian(2015)]{sormani+2015} Sormani, M.~C. \& Magorrian, J.\ 2015, \mnras, 446, 4186. %doi:10.1093/mnras/stu2316

\bibitem[Su et al.(2010)]{su+2010} Su, M., Slatyer, T.~R., \& Finkbeiner, D.~P.\ 2010, \apj, 724, 1044.% doi:10.1088/0004-637X/724/2/1044

\end{thebibliography}
\end{document}